\documentclass[journal,hideappendix]{vgtc}        

\usepackage[svgnames, dvipsnames]{xcolor}
\usepackage{amsmath}
\usepackage[framemethod=tikz]{mdframed}
\usepackage[numbers]{natbib}

\onlineid{0}

\vgtccategory{Research}

\vgtcpapertype{application/design study}

\title{\sysname{}: A Grammar for Visualizing Statistical Model Checks}

\author{Ziyang Guo, Alex Kale, Matthew Kay, Jessica Hullman}
\authorfooter{
\item
 Ziyang Guo is with Northwestern University. E-mail: ziyang.guo@northwestern.edu.
\item
 Alex Kale is with University of Chicago. E-mail: kalea@uchicago.edu.
\item
 Matthew Kay is with Northwestern University. E-mail: mjskay@northwestern.edu.
\item
 Jessica Hullman is with Northwestern University. E-mail: jhullman@northwestern.edu.
}

\usepackage{enumitem}
\usepackage{makecell}
\usepackage{annotate-equations}
\usepackage[normalem]{ulem}

\newcommand{\responsevar}{\textrm{y}}
\newcommand{\inputvar}{\textrm{x}}

\newcommand{\parameter}{\theta}
\newcommand{\pushforward}{\tau}
\newcommand{\linkfunction}{l}

\newcommand{\quantity}{q}

\definecolor{componentcolor}{RGB}{11, 129, 168}

\newcommand{\samplingspecification}[0]{\texttt{Sampling\_spec}}
\newcommand{\datatransformation}[0]{\texttt{Data\_transformation}}
\newcommand{\visualrepresentation}[0]{\texttt{Visual\_representation}}
\newcommand{\comparativelayout}[0]{\texttt{Comparative\_layout}}

\newcommand{\new}[1]{#1}

\newcommand{\sysname}{\texttt{VMC}}

\abstract{%
  Visualizations play a critical role in validating and improving statistical models. However, the design space of model check visualizations is not well understood, making it difficult for authors to explore and specify effective graphical model checks. \sysname{} defines a model check visualization using four components: (1) samples of distributions of checkable quantities generated from the model, including predictive distributions for new data and distributions of model parameters; (2) transformations on observed data to facilitate comparison; (3) visual representations of distributions; and (4) layouts to facilitate comparing model samples and observed data. We contribute an implementation of \sysname{} as an R package. We validate \sysname{} by reproducing a set of canonical model check examples, and show how using \sysname{} to generate model checks reduces the edit distance between visualizations relative to existing visualization toolkits. The findings of an interview study with three expert modelers who used \sysname{} highlight challenges and opportunities for encouraging exploration of correct, effective model check visualizations.
}

\keywords{Model checking and evaluation; Uncertainty visualization; Grammar of Graphics}

\teaser{
  \centering
  \includegraphics[width=\textwidth]{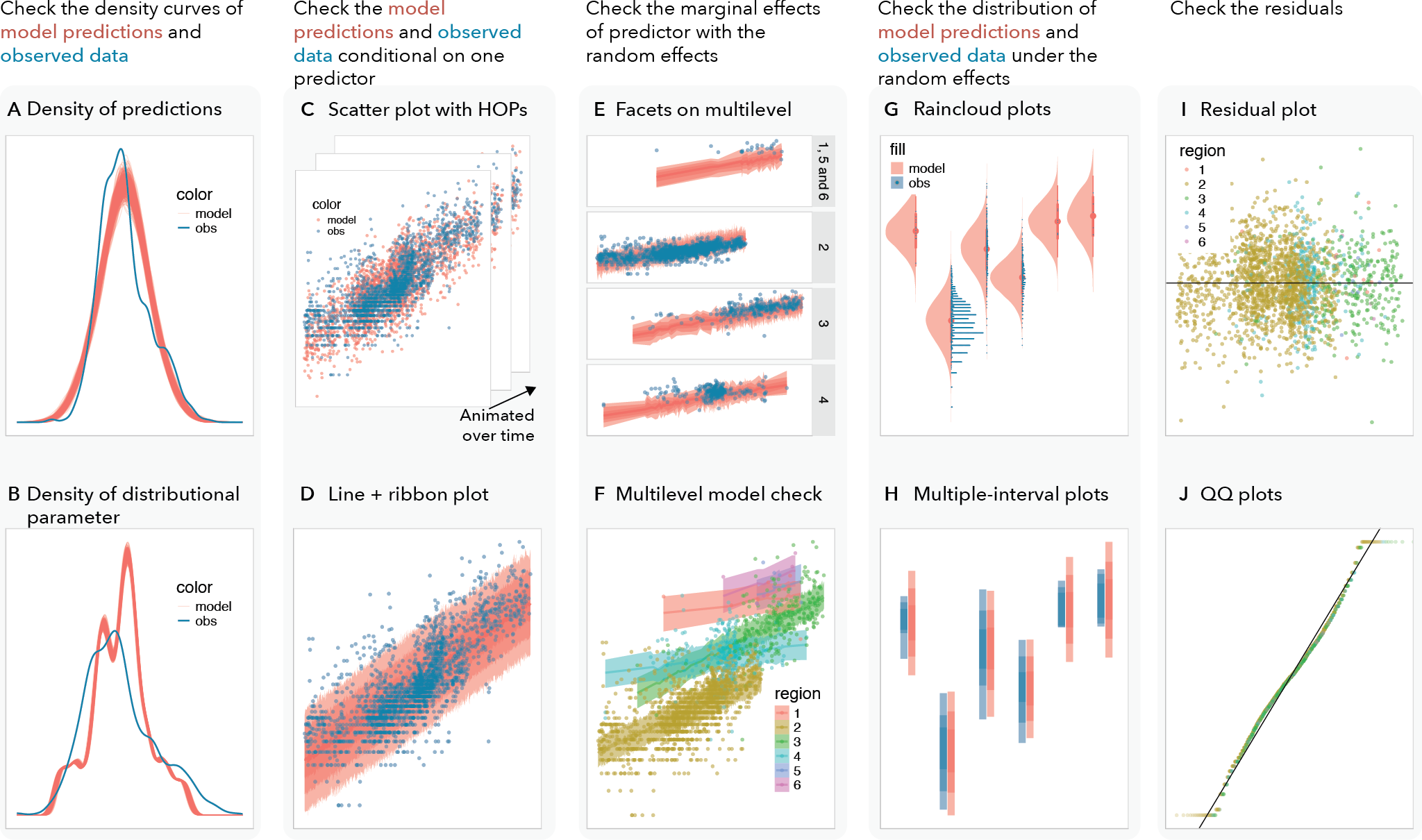}
  \caption{
  \new{
  Example model check visualizations authored with \sysname{}, using data from \cite{shaddick2018data}. From left to right: checks on the density curves of the distributions of model predictions and observed data
  from (A) response variable to (B) distributional parameter; follow-up checks conditional on the quantitative predictor, where \sysname{} is used to specify (C) Hypothetical Outcome Plots and (D) a line + ribbon plot; (E) a facet check stratifying the random effects and (F) a multilevel check; more checks for the random effects specified by \sysname{}, including (G) raincloud plots and (H) multiple-interval plots; and residual checks specified by \sysname{}, including (I) residual plots revealing the heteroskedasticity of the model
  and (J) Q-Q plots, validating the normality of residuals.
  }
  }
  \label{fig:teaser}
}

\graphicspath{{figs/}{figures/}{pictures/}{images/}{./}} 

\usepackage{tabu}                      
\usepackage{booktabs}                  
\usepackage{lipsum}                    
\usepackage{mwe}                       

\usepackage{mathptmx}                  

\begin{document}


\firstsection{Introduction}

\maketitle


Visualizations play a critical yet undervalued role for checking expectations in statistical modeling workflow.
A rigorous statistical analysis often includes numerical or graphical checks of how well the model fits the data and performs for the purpose for which it is developed~\cite{gelman1995bayesian, gabry2017visualization, gelman2004exploratory}. 
Experienced statisticians view numerical and graphical checks as serving complementary roles, and often use them in tandem.
For example, it is well-known that while a high R-squared value in a regression model can suggest a good fit, a residual plot might reveal heteroskedasticity or other issues not evident from the R-squared value alone~\cite{performance}. Model check visualizations also play a powerful role in communicating and helping validate visually-based inferences~\cite{gelman2004exploratory,hullman2019authors}, which can be strengthened by integrating functionality for generating model check visualizations in exploratory visual data analysis tools~\cite{Hullman2021Designing,kale2023evm}.

Graphical model checks (\textit{model checks} hereafter) extend far beyond well-known examples like residual plots, and can take many different forms. 
For example, they can juxtapose the model predictions and observed data side by side.
They can directly encode the difference between the two such as in residuals (\autoref{fig:teaser}I, J).
The visual representations used in model check visualizations also vary widely.
They can use different mark types to represent the model and observed data, such as raincloud plot in \autoref{fig:teaser}G.
They can be animated to reveal the uncertainty in model distributions, such as hypothetical outcome plots in \autoref{fig:teaser}C.


A lack of graphical tools for composing or exploring visualizations for model checking restricts the potential for these visualizations to improve statistical practices.
Expert modelers tend to rely on their previous experience to guide their design choices.
For example, some analysts may only think of residual plots to check assumptions of heteroskedasticity or density estimates to check for systematic discrepancies between model predictions and observed data~\cite{gelman1995bayesian, gabry2017visualization}.
Others may gravitate toward other views based on their training (e.g., Bayesian workflow~\cite{gelman2020bayesian}).
Reliance on prior experience alone can restrict exploration of model check visualization designs and employment of new techniques for displaying uncertainty~\cite{kay2023ggdist} or comparative layouts~\cite{gleicher2011visual}.

One way to scale effective visual model checking to more scenarios is to make it easier to generate a space of possible model check visualizations in statistical workflow.
Most current methods for specifying model checks either are too low-level to effectively support high-level tasks in model checks 
or have limited flexibility to explore diverse design choices.
For example, the Grammar of Graphics (e.g., \texttt{ggplot2}~\cite{ggplot2} and \texttt{Vega-Lite}~\cite{satyanarayan2016vega}) specify a visualization as a set of layers comprised of data, geometries, aesthetics, statistical transformations,  scales, and guides. 
However, low-level representations can lead to problems like representational \textit{viscosity}, \textit{hidden dependencies}, and \textit{error-proneness}~\cite{blackwell2001cognitive} when applied to designing model check visualizations.
For example, switching from a posterior predictive check on a density estimate (Fig.~\ref{fig:teaser}A) to a residual plot (Fig.~\ref{fig:teaser}I) using \texttt{ggplot2} requires changes to data, aesthetics, scales and geometry.
On the other hand, task-oriented approaches like \texttt{bayesplot}~\cite{gabry2017visualization, bayesplot} and \texttt{performance}~\cite{performance} only support fixed visualizations for specific tasks, 
discouraging further exploration.
This trade-off between simplicity and expressiveness motivates careful development of appropriate abstractions for model check visualization.
We contribute \sysname{}, a high-level declarative grammar for generating model check visualizations.
\sysname{} categorizes design choices in model check visualizations via four components: \textit{sampling specification}, \textit{data transformation}, \textit{visual representation(s)}, and \textit{comparative layout}.
\sysname{} improves the state-of-the-art in graphical model check specification in two ways: (1) it allows users to explore a wide range of model checks through relatively small changes to a specification as opposed to more substantial code restructuring, and 
(2) it simplifies the specification of model checks by defining a small number of semantically-meaningful design components tailored to model checking.

We implement \sysname{} as a package in the popular statistical programming language R.
By reproducing a range of model check techniques that occur in the statistical literature \cite{gelman1995bayesian}, we demonstrate how (1) the design space specified by the components of \sysname{} covers canonical examples of model checks used in statistics and (2) the abstractions employed by \sysname{} can reduce edit distance between model check visualization specifications compared to existing general purpose and specialized graphics libraries, e.g., \texttt{ggplot2} and \texttt{bayesplot}.
We further contribute the results of a semi-structured interview study in which three expert modelers applied \sysname{} to a model of their choice. We find that \sysname{} (1) aligns with participants' understanding and practice of model checks and (2) encourages exploration of new model checks.

\section{Related Work}
\label{sec:background}

\subsection{Visualizations as Model Checks}

Model checking is a fundamental part of statistical workflow~\cite{gelman1995bayesian, box1980sampling, blei2014build}.
Among forms of checks, visualizations that enable graphical checking has long been viewed as an important complement to quantitative statistical checks~\cite{buja2009statistical,gabry2017visualization,tukey1977exploratory}.
For example, modelers use residual plots to check model assumptions like heteroskedasticity~\cite{box1980sampling} and scatter plots to check for multimodality~\cite{ameijeiras2019mode}.
Gabry et al.~\cite{gabry2017visualization} summarizes the usage of graphical checks in each step of the Bayesian statistical workflow and implements these canonical checking examples in an R graphical tool, \texttt{bayesplot}.

Model checks are designed to facilitate the comparison between observed data and predictions under the model. The best model check visualizations are said to reduce this task to detecting deviation from symmetry, for which the human visual system is well-optimized~\cite{gelman2004exploratory}.
Because model checks are inherently visual, they are hard to theorize, though statisticians have identified properties expected of a good model check.
For example, Li et al.~\cite{li2022calibrated} suggest a model check should be well-calibrated, computationally efficient, and easy-to-use.
We develop \sysname{} to address obstructions to the last property when it comes to existing libraries for constructing model checks.

More broadly, some prior work has explored the idea of strengthening the connection between visualization and statistical model checking. 
Early work in statistical graphics emphasized the value of trellis plots for checking for main effects and interactions between variables in plots of observed data~\cite{becker1996visual}.
Gelman proposed a theoretical formulation of visualizations as model checks to help bridge seemingly opposed traditions of exploratory and confirmatory data analysis~\cite{gelman2003bayesian, gelman2004exploratory}. 
Other authors have similarly proposed an analogy of a visualization as a statistical test, and contributed graphical tools like the line-up to facilitate graphical statistical inference~\cite{wickham2010graphical, hofmann2012graphical, majumder2013validation}. 
Hullman and Gelman compare statistical perspectives on the analogy between visualizations and statistical modeling, likening the examination of visualization to posterior predictive checking in a Bayesian workflow and deriving design implications of this view for interactive visual analysis software \cite{Hullman2021Designing, Hullman2021Challenges}.
Kale et al. \cite{kale2023evm} realize a version of this vision in EVM, a Tableau-style visualization tool that incorporates lightweight specification and graphical checking of models to help users express and check their provisional interpretations of data in exploratory visual analysis.
While generatlly associated with traditional statistical modeling, visual model checking also plays a role in machine learning pipelines, where the goals of a VIS4ML system often hinge on the ability of the system user to identify forms of model misfit~\cite{subramonyam2023we}. The diversity of applications for model checking emphasize the need for tools and theory to be agnostic to the specific goals and form of the model. 
Despite its importance as a statistical tool, there is little practical guidance on how to evaluate or construct model check visualizations \cite{Hullman2021Designing,muller2022forgetting}.

\subsubsection{Visualizations for comparison}

Model checks are a particular type of visualization-aided comparison.
Gleicher et al. \cite{gleicher2011visual} discuss the challenges in designing visualizations for aiding comparison and propose a taxonomy for comparison techniques in visualizations, which informs the comparative layout component of our grammar.
Elsewhere, canonical work in visualization highlights the value of faceting data for checking implicit models~\cite{becker1996visual}.
Our work extends understanding of visual comparison by focusing specifically on the design space for 
facilitating comparison between model predictions and observed data. 

\subsection{Visualization Grammars}
Visualization grammars can provide a more rigorous basis for generating visualizations over chart-type typologies, as well as a principled way of generating a search space of visualizations for visualization recommenders to act on. 
For example, the \textit{Grammar of Graphics} (implemented by \texttt{ggplot2}) \cite{wilkinson2012grammar} specification uses layered components to encode visual representation layers, scales for aesthetic attributes, a coordinate system to map the position on the plot, and facets to partition data into subsets.
Some visualization grammars target more specific domains.
For example, \texttt{ggdist}~\cite{kay2023ggdist} and \texttt{PGoG}~\cite{pu2020probabilistic} provide a particular formalism for specifying uncertainty visualizations as an extension to \texttt{ggplot2}, and \textit{IEMA}~\cite{baniecki2023grammar} provides a formalism to specifying explanatory model analysis.

However, existing visualization grammars are not optimized for specifying model checks.
Constructing model checks usually involves techniques like extracting samples from the model, visualizing the samples in visual representation, and comparing model predictions and observed data in layouts.
However, as discussed by literature~\cite{pu2023data}, existing visualization grammars such as \texttt{ggplot2}, can lead to specific errors such as mismatched data and visualization due to the lack of tight couplings between data transformation and visualization code and the difficulty of keeping these in sync.
Model check libraries like \textit{bayesplot} \cite{gabry2017visualization} provide users a quick API to define a model check visualization by generating fixed plots for certain model check tasks, e.g., a density plot for posterior predictive checks and residual plots for checking heteroskedasticity. 
However, they are not meant to provide a systematic understanding of what constitutes a model check that users can rely on to explore or invent designs.




\section{Design Objectives}
Drawing on the aforementioned prior work on model checking workflow, 
we derived four design objectives for a model check visualization grammar.

\begin{enumerate}[wide, labelindent=0pt,nosep,label={\bf O{{\arabic*}}}]
\item \label{req:expressive} \textbf{Expressiveness}.
To scale the benefits of model checking to many scenarios, a model check visualization grammar should be able to express as many model check design strategies as possible, incorporating different visualization techniques.
One way to enable visualization flexibility is to componentize data preparation and visual specification separately.
Data preparation components should support varying methods to access the model and observed data.
The visual aspects of a grammar should control how to render the model and observed data (e.g., represent the model by lines and observed data by points) and how to lay them out to facilitate comparison (e.g., aligning the plot for the model and the plot for the observed data spatially or overlaying them in one same coordinate system).

\item \label{req:correctness} \textbf{Correctness}
Following other task-oriented visualization grammars~\cite{pu2020probabilistic}, a model check visualization grammar should support generation of ``correct'' examples only. 
In the context of  model check visualization, correctness corresponds to specifications of visualization that are a function of both model predictions and observed data.  
Prior work on visualization grammars suggest that one way to avoid generation of incorrect examples is for the visualization grammar to tightly couple the data and visualization specifications, as mismatch between them can lead to errors~\cite{pu2023data}. 
Maintaining this synchronization can be particularly complex in the case of model check visualizations, which requires keeping two different data pipelines (for observed data and model predictions) and their corresponding representations in sync.


\item \label{req:closeness} \textbf{Closeness to the experts' language}
A model check visualization grammar should use abstractions that resemble language experts use in specifying model checks, including both to describe the relevant data processes and to specify the visualization.
Ideally, a grammar will enable specifying the data corresponding to a desired model check visualization at a high level corresponding to major distinctions between methods experts care about but with sufficient control over crucial details like sampling methods that experts may care about.
For visual choices, possible specifications should correspond to the space of combinations of meaningful variations in design choices observed in expert model checking workflow.

\item \label{req:exploration} \textbf{Exploration}.
Following other grammars that facilitate exploration, a grammar should be component reusable~\cite{heer2005prefuse}, coherent and systematic~\cite{cox1978some}.
For component reusability, the components of the specification of a model check visualization grammar should be independent of one another such that changing one component does not necessarily require changing others.
To be coherent and systematic, the specification of a model check visualization grammar should be hierarchical and the components should control different parts of model check visualization in a systematic way.

\end{enumerate}

\section{\sysname{}: Grammar for Model Checks}
\label{sec:grammar}



\begin{figure*}
    \centering
    \includegraphics{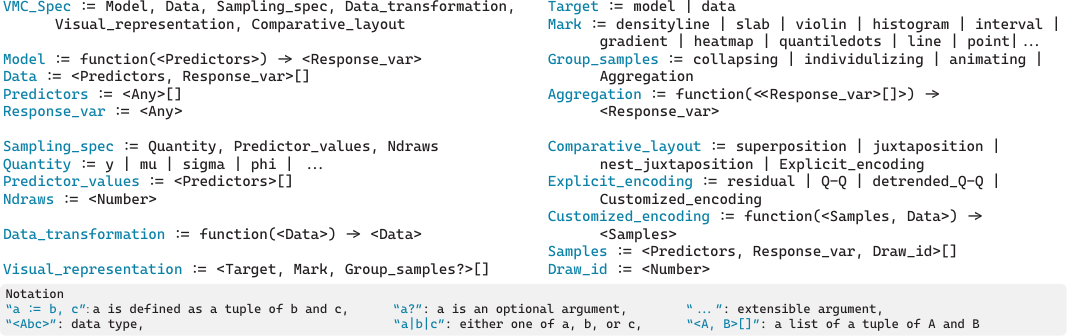}
    \caption{The formal specification of \sysname{} including the four main components: \samplingspecification{}, \datatransformation{}, \visualrepresentation{} and \comparativelayout{}.
    }
    \label{fig:notation}
    \vspace{-6mm}
\end{figure*}

\sysname{} is a declarative grammar designed to support concisely and flexibly specifying model check visualizations.
\sysname{} allows users to specify the data preparation and visual design of model check visualizations independently (\ref{req:expressive} and \ref{req:exploration}) with tailored components for common model checking tasks (\ref{req:closeness}).
A single \sysname{} specification defines how to generate samples from the model (\samplingspecification{}), transform the observed data (\datatransformation{}), visualize the model samples and the observed data (\visualrepresentation{}) and set up comparison between the model and the observed data (\comparativelayout{}).
\sysname{} takes an input \textit{model} and a set of \textit{observed data} to be compared. 
The formal specification of \sysname{} is shown in \autoref{fig:notation}.


\subsection{Sampling Specification}
\label{sec:distribution}

The first step to specifying a model check visualization is to define a way to access the fitted model.
We build the \samplingspecification{} component of \sysname{} to control how to sample from the model.
We designed \sysname{}'s \samplingspecification{} to be robust to model types and various sampling methods in model checks to fulfill the objective of expressiveness (\ref{req:expressive}). 
At the meantime, it should also have arguments to succinctly express conceptually distinct sample generation options to maintain closeness to experts' language (\ref{req:closeness}) and facilitate exploration (\ref{req:exploration}).
The \samplingspecification{} component of \sysname{} covers varieties of sampling methods with three arguments: (1) the quantity to be generated from the model, (2) the data set of predictor values, and (3) the number of draws.

There are many different model-defined quantities that can be useful to check in model check visualizations ~\cite{ryan2008modern, gelman2003bayesian, velleman1981efficient}. 
For example, in Bayesian data analysis~\cite{gelman2003bayesian}, modelers frequently sample from the posterior distribution of the model parameters in addition to drawing from the posterior predictive distribution.
The quantities that are being checked also vary with the choice of link functions.
For example, modelers often check the link-scale predictions of the logarithm of the odds ratio in logistic regression models~\cite{insight}.

\samplingspecification{} formalizes the quantities within a statistical model by combining link functions with the response variable and distributional parameters, if applicable.
For example, in the model shown in \autoref{eq:model}, the quantities included by \samplingspecification{} are $\responsevar$, $\linkfunction(\responsevar)$, $\pushforward_1, \ldots, \pushforward_M$ and $\linkfunction_1(\pushforward_1), \ldots, \linkfunction_M(\pushforward_M)$.
The definition of quantities within \samplingspecification{} is designed to balance expressiveness and redundancy; we aimed for \samplingspecification{} to express as many model quantities as possible while ensuring that it only includes quantities that are \textit{checkable}.

A \textit{checkable} quantity is one for which there exists a method for comparing it with observed data.
Distributional parameters often correspond to statistics computed on the observed data, such as $\mu$ in a Gaussian distribution corresponding to the mean of the observed data and $\sigma$ corresponding to the standard deviation of observed data.
There will however be cases where we cannot check arbitrary quantities generated under the model against observed data.
For example, consider a scenario where an author wants to check $\alpha$ in the model:
$y \sim  \mathit{Normal}(\mu, \phi), \mu =  \alpha + \beta x$.
To compare with $\alpha$ in the model
requires prior information on the true data generating process of the response variable $\responsevar$. However, uncertainty about the true data generating process is typically why explanatory models are used in the first place.

\vspace{10mm}
\begin{align} 
\label{eq:model}
\begin{split}
    \eqnmarkbox[BrickRed]{linky1}{\linkfunction}(
    \eqnmarkbox[NavyBlue]{y1}{\responsevar} )
    &\sim p( 
    \eqnmarkbox[BrickRed]{linky2}{\linkfunction}(
    \eqnmarkbox[NavyBlue]{y2}{\responsevar} )
    | \eqnmarkbox[OliveGreen]{t}{\pushforward_1, \ldots, \pushforward_m, \ldots, \pushforward_M}) \\
    \eqnmarkbox[BrickRed]{l}{\linkfunction_m}
    (\pushforward_m) &= f_m(
    \eqnmarkbox[RubineRed]{theta1}{\parameter}, 
    \eqnmarkbox[RedViolet]{x1}{\inputvar})
\end{split}
\end{align}
\annotatetwo[yshift=2.5em]{above}{y1}{y2}{response variable}
\annotatetwo[yshift=1em]{above}{linky1}{linky2}{link function}
\annotate[yshift=-1em]{below, label below}{theta1}{model parameters}
\annotate[yshift=-2.5em]{below, label below}{x1}{independent variables}
\annotate[yshift=4em]{above,left}{t}{distributional parameters}
\annotate[yshift=-1em]{below,left}{l}{link function}
\vspace{7mm}

\subsection{Data Transformation}
\label{sec:transformation}
Given the varieties of data transformations that can be applied in model check visualizations~\cite{gelman1996posterior, gelman2003bayesian, gabry2017visualization}, we define \datatransformation{} as a function that takes in a data set of observed data and outputs a data set with the same format as the input data.
For example, instead of the original empirical distribution of the observed data, modelers may want to compare the model to certain statistics computed on the observed data, e.g., mean or median~\cite{gelman2003bayesian}.
Some transformations on the observed data are meant to help test a hypothesis, e.g., whether the model's predictive distribution captures the mean or median of the observed data. Others are performed to ensure comparability of the model quantity and observed data.
For example, when model samples are generated from the distribution of the $\sigma$ parameter in a Gaussian regression model, the model quantity is on a variance scale.
In this case, the observed data should first be transformed to its standard deviation.

\subsection{Visual Representation}

Effective visual representations---those that can enable the analyst to identify discrepancies between model predictions and observed data---are critical to the utility of model checking in a statistical workflow.
The third component of \sysname{}, \visualrepresentation{}, is used to specify the visual representations of the model and observed data separately.
It enables specifying a list of visual representations to compose a model check visualization. 
Authors can specify more than one visual representation for model predictions and observed data. 
For each visual representation for model predictions, in addition to the mark type, they can specify the method to use to group the samples in that representation.

\subsubsection{Choice of mark type}

Drawing on research on visualizing distributional information \cite{kay2023ggdist, pu2020probabilistic}, we observe that visualizations of uncertainty can be distinguished based on whether they use the area or extent of a mark to encode distributional information (e.g., a confidence interval or range), a visual variable such as color to encode probability, or a discrete (countable) representation in which distinct marks are used to visually compose a distribution.
 \sysname{} defines three categories for marks:

\vspace{-5mm}
\begin{align*}
    \textsc{extent}: & \ \textit{densityline}, \textit{slab} , \textit{violin} , \textit{histogram} , \textit{interval} , ... \\
    \textsc{visual variable}: & \ \textit{gradient} , \textit{heatmap} , ... \\
    \textsc{countable}: & \ \textit{quantiledots} , \textit{line} , \textit{point} , ...
\end{align*}
\vspace{-4mm}

In practice, the choice of mark type for visually representing distribution in model checks may be driven by the audience of the model checks and specific model checking tasks.
For example, if the audience is broader than just experts,
the authors 
may opt for a mark type from the countable category, e.g., quantile dotplots \cite{kay2016ish}, since countable visual representations appear to outperform continuous equivalents like error bars or densities for some tasks for lay audiences \cite{fernandes2018uncertainty,hullman2015hypothetical,hullman2017imagining,padilla2020uncertainty}. (The meaning of \textit{collapsing} and \textit{individualizing} in these specifications is described in section~\ref{sec:groupingsamples}.)

\vspace{-5mm}
\begin{equation*}
\begin{aligned}
    & \visualrepresentation{} \leftarrow \\
    & \quad [(model, \textit{quantiledots}, \\
    & \quad \quad collapsing), \\
    & \quad \ \ (data, \textit{quantiledots})]
\end{aligned}
\quad
\vcenter{\hbox{\includegraphics[width=4cm,height=2cm]{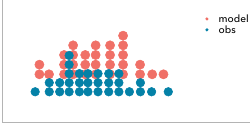}}}
\end{equation*}

The choice of mark type can also depend on the specific task the model check is designed for.
For example, the author may prefer a mark type based on intervals if they want to check calibration (i.e., whether the model predictions capture the expectation of observed data) or heteroskedasticity (i.e., whether the variance of residuals is constant). 

\vspace{-5mm}
\begin{equation*}
\begin{aligned}
    & \visualrepresentation{} \leftarrow \\
    & \quad [(model, \textit{interval}, \\
    & \quad \quad collapsing)), \\
    & \quad \ \ (data, \textit{interval})]
\end{aligned}
\quad
\vcenter{\hbox{\includegraphics[width=4cm,height=2cm]{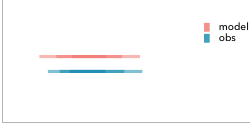}}}
\end{equation*}

If they want to check the shape of the distribution, they may prefer densities.

\vspace{-5mm}
\begin{equation*}
\begin{aligned}
    & \visualrepresentation{} \leftarrow \\
    & \quad [(model, \textit{densityline}, \\
    & \quad \quad individualizing)), \\
    & \quad \ \ (data, \textit{densityline})]
\end{aligned}
\quad
\vcenter{\hbox{\includegraphics[width=4cm,height=2cm]{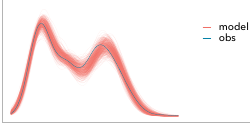}}}
\end{equation*}

They may also want to show more complex information about the model predictions and observed data, leading them to adopt hybrid mark types.
For example, statisticians have used raincloud plots \cite{allen2019raincloud}, a combination of several chart types, to visualize the raw observed data, the distribution of the model predictions as density, and key summary statistics at the same time.
 
\vspace{-5mm}
\begin{equation*}
\begin{aligned}
    & \visualrepresentation{} \leftarrow \\
    & \quad [(model, \textit{slab}, \\
    & \quad \quad collapsing)), \\
    & \quad \ \ (model, \textit{interval}, \\
    & \quad \quad collapsing)), \\
    & \quad \ \ (data, \textit{quantiledots})]
\end{aligned}   
\quad
\vcenter{\hbox{\includegraphics[width=4cm,height=2cm]{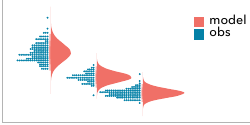}}}
\end{equation*}

\subsubsection{Grouping samples}
\label{sec:groupingsamples}
By sampling from the model, we yield multiple sets of model samples for visualizing.
Different approaches to grouping these samples prior to applying visual marks produce different visual formats that may differ in how well they support various tasks.
We summarise four ways of grouping samples into visual marks observed in model checking practice.

First, one common approach is to \textit{collapse} all samples into one visual mark.
Consider the example of a line ribbon plot in \autoref{fig:teaser}D: there is only one mark (line + ribbons for multiple confidence intervals), which is calculated from the data of all the samples.
Another way is to use \textit{individual} marks for each of the samples.
For example, in the posterior predictive check in \autoref{fig:teaser}A, each of the light red line represents the density estimate of one sample.
Hypothetical outcome plots (HOPs)~\cite{hullman2015hypothetical} animate one mark of one sample at each time frame, e.g., \autoref{fig:teaser}C.
Finally, sometimes modelers may want to plot the distribution of aggregate statistics of each sample (e.g., a histogram of sample means~\cite{gelman2003bayesian}). 

\visualrepresentation{} uses \texttt{Grouping\_samples} to control how to group the model samples into the visual marks.
\sysname{} supports grouping samples using four options: \textit{collapsing}, \textit{individualizing}, \textit{animating}, and \textit{aggregating}.
Let $\mathbf{\quantity}^{(i)} = [\quantity_1^{(i)}, \ldots, \quantity_n^{(i)}]$ be a sample from the model on a set of observed values of predictors, $\mathbf{x} = [x_1, \ldots, x_n]$, where $n$ is the number of instances.
There are $r$ samples total: $\mathbf{\quantity}^{(1)}, \ldots, \mathbf{\quantity}^{(r)}$.
Using \textit{collapsing} will visualize all samples in one visual mark, i.e., the visual mark takes as input $\mathbf{\quantity}^{(1)}, \ldots, \mathbf{\quantity}^{(r)}$.
Using \textit{individualizing} and \textit{animating} are both specified to visualize samples in separate visual marks: 
\textit{individualizing} overlays all marks in one coordinate system while \textit{animating} shows each visual mark in one time frame.
Thus, when using \textit{individualizing} or \textit{animating}, there are $r$ visual marks with each of them takes input as $\mathbf{\quantity}^{(i)}$.
Using \textit{aggregating} will aggregate statistics of each sample and then use one mark for the aggregated statistics.
Denoting the aggregating function as $\pi: \mathrm{R}^n \rightarrow \mathrm{R}$,
when using \textit{aggregating}, the visual mark takes as input $\pi(\mathbf{\quantity}^{(1)}), \ldots, $ $\pi(\mathbf{\quantity}^{(r)})$.

\begin{figure}
    \centering
    \includegraphics[width=\linewidth]{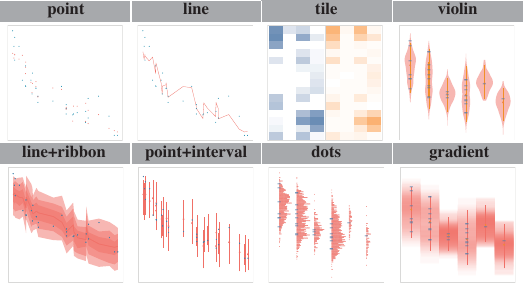}
    \caption{A subset of visual representations that \sysname{} can specify. 
    Not shown: Animation (HOPs) can be applied upon any.}
    \label{fig:uncertainty_representation}
    \vspace{-8mm}
\end{figure}
\subsection{Comparative layout}
\label{sec:layout}


The final necessary decision in visualizing a model check is to specify a comparative layout to facilitate checking the alignment of data observations and model predictions.
We summarise four classes of comparative layouts from the visual comparison literature~\cite{gleicher2011visual, becker1996visual, ondov2018face, jardine2019perceptual} : \textit{superposition}, \textit{juxtaposition}, \textit{nested juxtaposition}, and \textit{explicit encoding}.
Our grammar makes these options entities of the same type (comparative layout), which is notably different from the grammar of graphics, where researchers have noted a mismatch between author’s comparison tasks and Grammar of Graphics components~\cite{pu2023data}.
We represent the choices of comparative layouts directly in \comparativelayout, to better match model check visualization authors' tasks.
\textit{Juxtaposition} and \textit{superposition} do not require changes to the visual representations of model predictions and observed data, but organize the layout to make them spatially aligned \cite{matlen2020spatial}. 

\vspace{-5mm}
\begin{equation*}
\begin{aligned}
    & \visualrepresentation{} \leftarrow \\
    & \quad [(model, \textit{lineribbon}, \\
    & \quad \quad collapsing), \\
    & \quad \ \ (data, \textit{lineribbon})] \\
    & \comparativelayout{} \leftarrow \\
    & \quad \textit{juxtaposition} 
\end{aligned}
\quad
\vcenter{\hbox{\includegraphics[width=4cm,height=2cm]{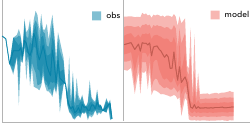}}}
\end{equation*}

\vspace{-5mm}
\begin{equation*}
\begin{aligned}
    & \visualrepresentation{} \leftarrow \\
    & \quad [(model, \textit{histagram}, \\
    & \quad \quad collapsing), \\
    & \quad \ \ (data, \textit{histagram})] \\
    & \comparativelayout{} \leftarrow \\
    & \quad \textit{superposition} 
\end{aligned}
\quad
\vcenter{\hbox{\includegraphics[width=4cm,height=2cm]{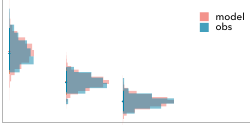}}}
\end{equation*}
\textit{Nested juxtaposition} is a variant of juxtaposition, where the distributions of model predictions and observed data are placed side by side but within the coordinate system of the plot.
\vspace{-2mm}
\begin{equation*}
\begin{aligned}
    & \visualrepresentation{} \leftarrow \\
    & \quad [(model, \textit{pointinterval}, \\
    & \quad \quad collapsing), \\
    & \quad \ \ (data, \textit{pointinterval})] \\
    & \comparativelayout{} \leftarrow \\
    & \quad \textit{nested\_juxtaposition} 
\end{aligned}
\quad
\vcenter{\hbox{\includegraphics[width=4cm,height=2cm]{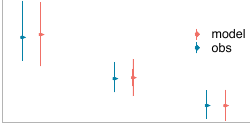}}}
\end{equation*}
To use \textit{explicit encoding} requires calculating a transformation on the model predictions and observed data \cite{gleicher2011visual}.
Explicit encoding is often used to facilitate checking specific assumptions of a model, such as heteroskedasticity, linearity, and normality of errors (Fig. \ref{fig:teaser}I and J).
\vspace{-2mm}
\begin{equation*}
\begin{aligned}
    & \visualrepresentation{} \leftarrow \\
    & \quad [(model, \textit{lineribbon}, \\
    & \quad \quad collapsing)] \\
    & \comparativelayout{} \leftarrow \\
    & \quad \texttt{Explicit\_encoding} \\
    & \texttt{Explicit\_encoding} \leftarrow \\
    & \quad \textit{residual} 
\end{aligned}
\quad
\vcenter{\hbox{\includegraphics[width=4cm,height=2cm]{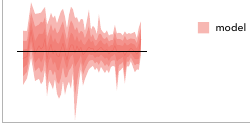}}}
\end{equation*}

An effective comparative layout renders differences between model predictions and observed data visually salient.
Which layout works best will depend on the specific circumstances.
Superposition facilitates comparisons since the visual representations of model predictions and observed data are plotted along the same axes, but also introduces overlap that may result in loss of information on the part of the viewer.
For example, when using a visual representation that encodes uncertainty information with opacity, a superposition layout will blend the opacity for model predictions and observed data.
Juxtaposition avoids overlap but requires visuo-spatial working memory and eye movements to enable comparison.
Nested juxtaposition is helpful for detecting local differences because it places the information being compared close in space on a common, aligned scale.
It can avoid the shortcomings of superposition and juxtaposition, but sometimes requires more work to create. 
For example, creating a nested juxtaposition for a model check that is conditional on a continuous variable can be complicated; the author might need to cut the continuous axis into several segments and intermittently place the segments of model predictions and observed data.
Explicit encoding presents the comparison directly, avoiding information loss, but also introduces abstraction by changing the unit. 
For example, Q-Q plots show the relationship between model predictions and observed data by plotting the quantiles of model predictions against the quantiles of observed data.

\section{A Walk-through Example: Air Quality}
\label{sec:walkthrough}
\new{
We use the data from Shaddick et al.~\cite{shaddick2018data} and models from Gabry et al.~\cite{gabry2017visualization} to demonstrate \sysname{}'s usage in a statistical workflow.
The focus is $\text{PM}_{2.5}$, representing exposure to air pollution from particulate matter measuring less than 2.5 $\mu$m in diameter, that is linked to a number of poor health outcomes.
Direct measurements of $\text{PM}_{2.5}$ rely on a sparse network of ground monitors with heterogeneous spatial coverage.
To obtain estimates of $\text{PM}_{2.5}$ concentration with high spatial resolution, direct measurements are supplemented with high resolution estimates from satellite data.
By calibrating the satellite estimations using the ground monitor data, the goal is to obtain high-resolution, high-precision estimates of $\text{PM}_{2.5}$.}
\new{We start with a simple linear regression:
\begin{align}
\vspace{-5mm}
\label{eq:example_beta_model_initial}
\begin{split}
    \mathrm{log}(\text{PM}_{2.5}) \sim & \mathit{Normal}(\mu, \sigma) \\
    \mu = & \alpha + \beta \mathrm{log}(\text{satellite})\\
    \mathrm{log}(\sigma) \sim & \mathit{Normal}(0, 1)
\end{split}
\vspace{-5mm}
\end{align}
After fitting the model, we use \sysname{} to check the model's predictions against the correlation between the observed $\text{PM}_{2.5}$ and satellite estimates.
We start by checking the overall distribution of the model predictions of $\text{PM}_{2.5}$ compared with the observed $\text{PM}_{2.5}$. 
We set the \texttt{Quantity} in \samplingspecification{} as \texttt{y}, which generates \autoref{fig:teaser}A.
With separate \samplingspecification{} and \visualrepresentation{} components in \sysname{}, we can also flexibly check on different posterior distributions (e.g., \autoref{fig:teaser}B) and use different visual representations of model distribution and observed distribution.
This initial model check visualization provides a good starting point by revealing the misalignment between the shape of the distribution of predicted $\text{PM}_{2.5}$ and observed $\text{PM}_{2.5}$, where skewness in the observed data is not captured by the model. }

\new{
Next we explore causes of this skewness to improve the initial model.
We look at the marginal effects of the predictor to $\text{PM}_{2.5}$.
\sysname{} supports this step with various visual formats.
First, we set \texttt{Mark} in \visualrepresentation{} as point for both the model distribution and observed distribution.
We use \texttt{Group\_samples} to animate the samples (Hypothetical Outcome plots \cite{hullman2015hypothetical}), which gives a view of what individual samples from the model distribution look like (\autoref{fig:teaser}C).
Using points reveals the original model predictions clearly but makes it hard to perceive the uncertainty directly, so we use a different \texttt{Mark}, line + ribbons (\autoref{fig:teaser}D) to display the uncertainty intervals. }

\begin{figure}
    \centering
    \includegraphics{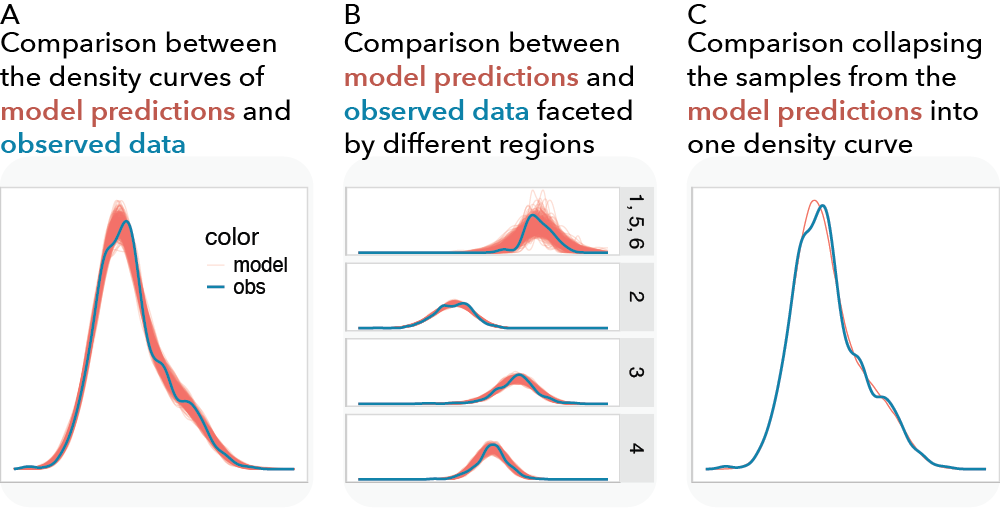}
    \caption{\new{The model check examples for updated model. 
    }}
    \label{fig:newmodel_usage_scenario}
    \vspace{-5mm}
\end{figure}

\new{
We continue to explore other visualizations by cycling through components of \sysname{}, including trying different \comparativelayout{}s to compare the model predictions and observed data such as juxtaposition, and trying different faceting options to explicitly show how observed data is clustered (\autoref{fig:teaser}E).
When faceting by the regions, the marginal effects of $\text{PM}_{2.5}$ to satellite estimates have noticeable clusters in different regions (\autoref{fig:teaser}E).
In response, we update the model by adding submodels of regions.
\begin{align}
\vspace{-5mm}
\label{eq:example_beta_model}
\begin{split}
    \mathrm{log}(\text{PM}_{2.5})_{\color{red}r} \sim & Normal(\mu_{\color{red}r}, \sigma) \\
    \mu_{\color{red}r} = & \alpha_{{\color{red}r}} + \beta_{{\color{red}r}} \mathrm{log}(\text{satellite})\\
    \sigma \sim & Normal(0, 1)
\end{split}
\vspace{-5mm}
\end{align}
After fitting the new model, we check it, 
first by generating a canonical check used for multilevel models (\autoref{fig:teaser}F).
We also check how the submodels perform by specifying model checks that are conditional on the regions (\autoref{fig:teaser}G and H).
To do this, first, we set the \texttt{Mark} to encode uncertainty, e.g., \texttt{halfeye} for model distribution and \texttt{dots} for observed distribution in a raincloud plot (\autoref{fig:teaser}G).
Then, we adjust \comparativelayout{} to \texttt{nest\_juxtaposition} to reduce visual overlap.
Finally, to verify that this change improved the model, we return to our initial model check where we identified skewness (\autoref{fig:newmodel_usage_scenario}A) and facet by regions this time.
We validate that the density curves of the model predictions converge to the curve of the observed data well, but wonder how the overall density of all samples looks compared with the observed data.
\sysname{}'s design of \texttt{Group\_sample} in \visualrepresentation{} supports flexible operations on model samples such as collapsing all the samples from the model prediction into one density curve (\autoref{fig:newmodel_usage_scenario}C).}

\new{
Next, we look more deeply into the assumptions of the model.
We first check the residuals.
By using \sysname{}, we can do this quickly with \texttt{Explicit\_encoding} in \comparativelayout{} component, where we specify a function encoding the comparison as the difference between model predictions and observed data (\autoref{fig:teaser}I).
Then we check the normality of the residuals by specifying a Q-Q plot (\autoref{fig:teaser}J).
Through \sysname{}, we only need to change the encoding function set in \texttt{Explicit\_encoding}.
Looking at the residual plot, we validate that the variance of residuals is constant along the conditional variable.
In the Q-Q plot, however, we observe further misalignment, where the residuals do not follow the standard normal distribution, motivating further changes to the model spec. The goal behind \sysname{} is to make it easier for the analyst to focus on the modeling task at hand in such scenarios, with the grammar facilitating less tedious construction of checks.}

\section{Implementation in R}



We implement \sysname{} in the R language as a package, provided in Supplemental Materials.
The specification of \sysname{} follows conventions popularized by the widely-used \texttt{ggplot2} package and outputs a \texttt{ggplot} specification to generate the model check plot.
To start specifying a model check plot, the user issues a statement called \texttt{mcplot()} and then follows it with statements specifying the four requisite components of \sysname{}: \texttt{mc\_draw()} for \samplingspecification{}; \texttt{mc\_observation\_transformation()} for \datatransformation{}; \texttt{mc\_model\_*()} and \texttt{mc\_obs\_*()} for \visualrepresentation (where \texttt{*} stands for different mark types, e.g., \texttt{mc\_model\_slab()} for using \textit{slab} for model predictions); and \texttt{mc\_layout\_*()} for \comparativelayout (where \texttt{*} stands for different layers, e.g., \texttt{mc\_model\_superposition()} for using \textit{superposition}).
The implementation of \sysname{} also supports other supplementary components, e.g., \texttt{mc\_condition\_on()} to define the conditional variable used in the check and \texttt{mc\_gglayer()} to append a \texttt{ggplot2} layer to the output specification.

\section{Reproducing Real-world Examples}

\begin{figure*}
    \centering
    \includegraphics[width=\textwidth]{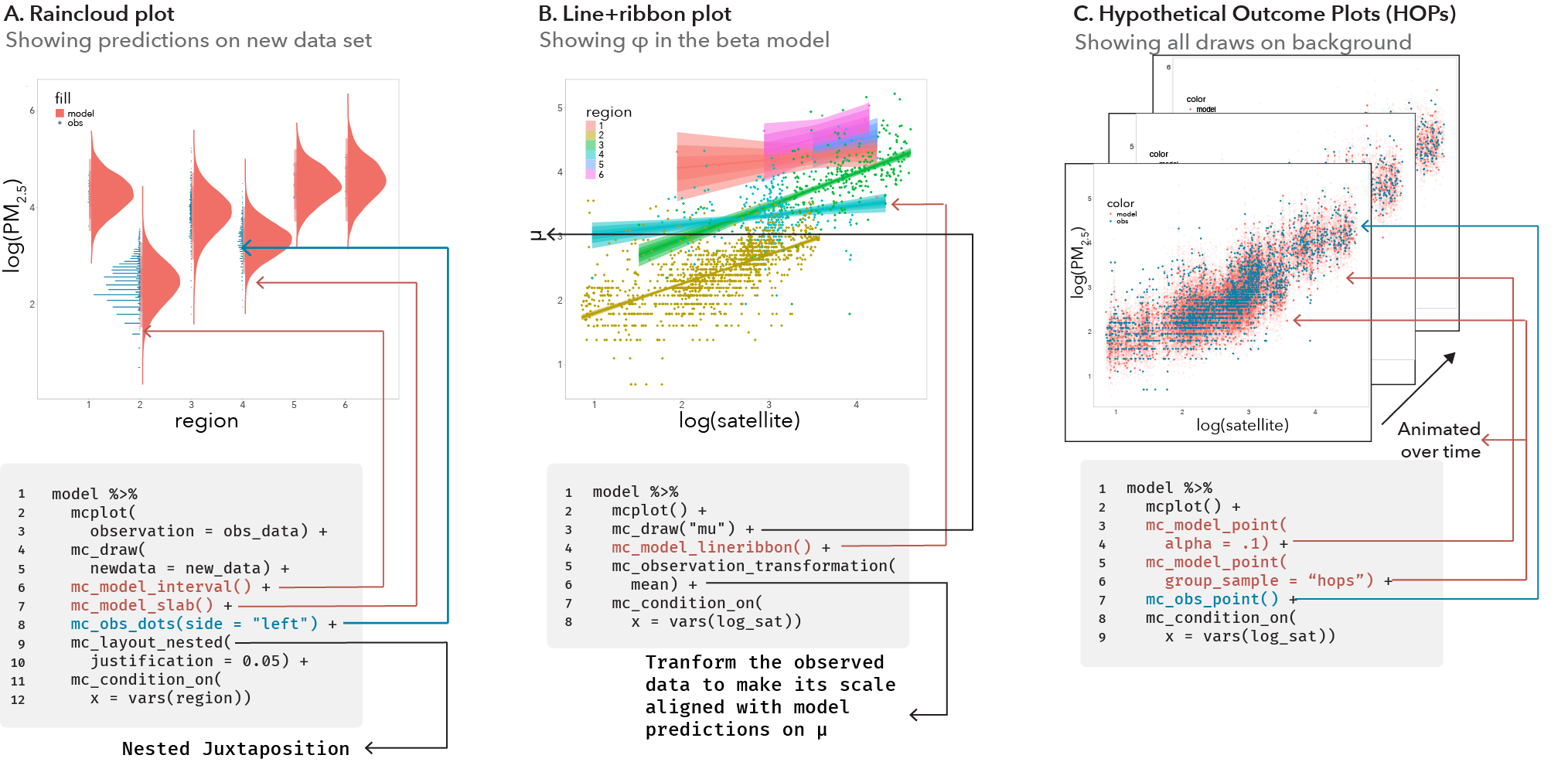}
    \caption{(A) Raincloud plots are constructed by two visual representation layers for model predictions and one for the observed data. (B) A line + ribbon plot is a combination of a line showing the mean of predictions and ribbons showing the prediction intervals. (C) HOPs uses animation to combine the samples from the model.}
    \label{fig:expressiveness}
    \vspace{-5mm}
\end{figure*}

To validate the expressiveness of \sysname{} as a visualization grammar, we collected and reproduced 27 model check visualization examples from statistics and visualization literature \cite{gelman1995bayesian, gelman2003bayesian, gelman2004exploratory, Hullman2021Designing, fernandes2018uncertainty, kale2023evm, yang2023swaying, yang2018correlation} that cover a description of model check techniques from a popular Bayesian statistics textbook \cite{gelman1995bayesian}\footnote{\new{We reproduce these examples using the implementation of \sysname{} in R.}}.
They highlight four model check techniques: \textit{replicated or external checks}, to make predictions under the fitted data or future data, \textit{choosing model quantities}, to check the aspect of the model we want, 
\textit{marginal predictive checks}, to check the joint predictive distribution, and \textit{residual checks}, to check the errors the model makes.
We evaluate the syntax design of \sysname{} by comparing it with \texttt{ggplot2} \cite{ggplot2} and \texttt{bayesplot} \cite{gabry2017visualization} (a task-oriented R package for model check of Bayesian models) while reproducing these examples.
We used an example Beta regression model in reproducing these model check visualizations
\footnote{See the full specification of the example Beta regression model and the reproduced examples with their corresponding R code (using \sysname{}, \texttt{ggplot2}, and \texttt{bayesplot} respectively) in Supplemental Materials}.
Across all the examples we implemented to compare APIs, there were an average of $30\%$ fewer lines of code with \sysname{} than with \texttt{ggplot2}.

\textbf{Replicated or external checks.}
\sysname{} supports replicated or external checks using the \texttt{Predictor\_values} argument in the \samplingspecification{} component.
\autoref{fig:teaser}A shows a replicated check generated by \sysname{}, which presents the kernel density estimate of the observed data (blue curve), with density estimates for samples from the posterior predictive distribution (orange, lighter lines).
\autoref{fig:expressiveness}A provides the full \sysname{} specification for an external check where the user is creating a model check with a ``raincloud'' plot.
The specifications of \sysname{} are more concise than their \texttt{ggplot2} and \texttt{bayesplot} counterparts.
With \sysname{}, users need only specify one argument to input a new data set to predict on.
In contrast, with \texttt{ggplot2} or \texttt{bayesplot}, since there is no grammar specification on how the data is generated from the model, users need to manually define the extract function using fitted or new data set and tidy the prediction's data format.

\textbf{Choosing model quantities.}
\sysname{}'s \samplingspecification{} component enables various model quantities to be presented in model checks. 
Consider the previous example of the kernel density estimates check.
We can specify the \texttt{Quantity} argument in \samplingspecification{} as $\mu$ (i.e., parameter $\mu$ in the beta regression model) (\autoref{fig:teaser}B).
\autoref{fig:expressiveness}B shows a more concrete example with code in R.
To align the scale of observed data with that of the model quantity $\mu$, we apply a transformation on the observed data.
The \sysname{} specification for choosing the model quantity is, once again, more concise than the corresponding \texttt{ggplot2} and \texttt{bayesplot} example.
When exploring the possible choices of model quantity, \sysname{} allows users to consider the quantities as one entity, rather than a data processing step prior to plotting as when using \texttt{ggplot2} and \texttt{bayesplot}.
This helps keep data processing and visualization consistent, potentially avoiding errors~\cite{pu2023data}.

\textbf{Marginal predictive checks}
\sysname{} supports marginal predictive checks through flexible 
specification of visual representations and comparative layouts.
For example, \autoref{fig:expressiveness}C shows how \sysname{} reproduces a canonical example of marginal checks introduced by Gelman \cite{gelman2004exploratory} and combined with HOPs by Hullman and Gelman \cite{Hullman2021Designing}. 
Besides conditioning on the independent variables on the x-axis, \sysname{} also enables 
grouping
marginal checks by color, rows and columns (i.e., faceting)~\autoref{fig:teaser}E and F.
To check the marginal effects in the model, \sysname{} requires a shorter \textit{edit distance} to switch between conditional variables compared with \texttt{ggplot2}.
In \sysname{} specification, the conditional variables are specified by a component separated from the visual representation components, so to change the conditional variable only requires changing one component.
This abstraction makes \sysname{} closer to experts' language (\ref{req:closeness}), reducing the complexity compared to \texttt{ggplot2}.


\textbf{Residual checks}
\sysname{} can specify various residual checks with various tasks.
For example, \autoref{fig:teaser}I checks the heteroskedasticity by showing the residual values with a scatter plot and \autoref{fig:teaser}J checks the normality of residuals with a Q-Q plot. 
Residual checks can be used to check other assumptions of the model, e.g., for a linear relationship between residuals and values of the response variable 
(see the example from
Gelman \cite{gelman2004exploratory} in Supplementary materials).
\sysname{} specification
separates the \comparativelayout{} from the other components, enabling users to
focus on the data transformations required 
for
the residual checks.
In \texttt{ggplot2}, however, the users need to consider the changes in geometries and aesthetics to incorporate with the changes in the data transformations.
While \texttt{bayesplot} supports residual checks directly, there 
is some overhead when switching
between different model checks, as 
code cannot easily be reused across different model checks.

\section{Observational Interview Study of \sysname}
\label{sec:evaluation}

To understand how data analysts specify visualizations and interact with \sysname{}, we conducted an interview-based study of its use by three experienced modelers. We aimed to answer the following research questions:

\begin{enumerate}[wide, labelindent=0pt,nosep,label={\bf RQ{{\arabic*}}}:]
\item \label{rq:expressiveness} Does the grammar capture core aspects of experienced analysts' understanding and practices of model check visualization?
\item \label{rq:received} What aspects of the grammar's construction do participants find most useful, and what challenges or trade-offs do they face in using it?
\item \label{rq:exploration} Can the grammar help experienced modelers explore novel (to them) model check visualizations, and what features encourage this?
\end{enumerate}

\subsection{Participants}

We recruited three experienced statistical modelers who were comfortable working with generative models via social media posts.
All three participants had obtained Ph.D. degrees in quantitative disciplines, and regularly conducted statistical analysis as part of their research. 
Two of three participants (P1 and P2) had more than 4 years of professional experience with statistical modeling, while the other one (P3) had over 2 years.
Two of three participants (P2 and P3) teach bachelor- and graduate-level statistics, including the practice of model checking.
All three participants reported having used \texttt{ggplot2} and \texttt{bayesplot} to generate the model check visualizations.
We compensated the participants with a $\$60$ Visa Gift card for a one hour session.

\subsection{Study Design}

Before the session, each participant filled out a background questionnaire. We asked them to bring a model from their current work to apply \sysname{} to.
The session started with the participant describing their typical graphical model checking practices, followed by describing the model they brought to the session, including data, modeling goals, and model fitting and selection process.
Participants then completed a a brief \sysname{} tutorial where they were asked to examine and run example code in an R notebook. These examples included 19 model check visualizations specified using \sysname{} on an example multilevel Gaussian model.

In the first part of the tutorial, the examples demonstrated variations supported by the components of \sysname{} by reproducing canonical model check visualizations in literature, including the examples from the statistical textbook~\cite{gelman2003bayesian}.
The second part demonstrated the expressiveness of \sysname{} by incorporating visualization techniques that have not been widely used in current model check practices.
For example, we included examples of using \sysname{} to display uncertainty information by animation (i.e., HOPs) and the nested juxtaposition comparative layout.
Participants were asked to experiment with new combinations of these components as they went through the tutorial, and to think aloud while navigating these tutorial examples. This portion of the study lasted approximately 20 minutes.
After the tutorial, participants were asked to create model check visualizations for the model they brought, for approximately 20 more minutes.  
As they used the grammar, we occasionally prompted them with questions about their perceptions of the grammar, including about whether the visualizations were useful to them and how they conceived of aspects of the grammar specification.
Each session ended with an exit interview (20 minutes) that asked participants to reflect on their process of using
\sysname{} and how they see \sysname{} interfacing with their typical workflow. \footnote{See Supplementary Materials for interview questions and the R document.}

\subsection{Results}

We summarize observations from participants' usage of \sysname{} and the exit interviews.  

\subsubsection{RQ1: Alignment with Participants' Model Check Understanding and Practice}

In stepping through examples we provided in the tutorial, participants commented on how they recognized visualizations they commonly used.
For example, P1 stated that ``\textit{this [posterior predictive check] is usually the first thing I do...}'' and ``\textit{this [Q-Q] plot is exact what I would do on my model}'', and P2 stated that ``\textit{this [line+ribbon plot] is very nice. I like this. This is something that I usually kind of do on my own}.''
All the participants noted that the options included in \sysname{} largely capture visualization variations that would interest them in practice.
For example, when going through the examples that checked distributional parameters in the model, P1 stated that \texttt{mc\_draw} and \texttt{mc\_observation\_transformation} are both interfaces that capture his practices of checking on different aspects of the model -- ``\textit{I like this. So I can change between all the parameters that the model estimates}''.

Moreover, two of the three participants (P1 and P2) found that \sysname{} also included some model check visualizations that they had never thought about before.
P1 remarked that the raincloud plot (\autoref{fig:expressiveness}A) displaying the uncertainty information in model predictions and observed data was new to him, saying \textit{``I like this [violin plot]. I’ve never done anything like this before.''}
The participants were also impressed by the animation over samples supported by the grouping sampling argument.
P1 said ``\textit{this [animation] is really cool. Initially I can’t see the [model] draws within the Q-Q plot but now it’s over the animation}''.

The \comparativelayout{} component in particular drew interest, leading to comments about how combining different layouts and visual representation could produce additional effective examples.
For example, P1 found that juxtaposition can avoid visual clutter in some cases.
He said ``\textit{I think I like this one [juxtaposition] better [than superposition], because it was pretty hard to see when the observed data was right on top of the the model [predictions]}''.
When P2 saw our examples using nested juxtaposition, he was pleasantly surprised by its alternative to superposition--``\textit{this [nested juxtaposition] is great. It separates those bars [confidence intervals] from one another, but doesn't overlay them together}.''

Participants' perspectives on the direct comparison \sysname{} encouraged between the model and the observed data in visualizations at times varied from that assumed in developing the grammar.
For example, when adopting \sysname{} to their own model, P2 created a model check visualization where the model quantity and the quantity in the observed data were not perfectly aligned. The model samples were generated from the posterior predictive distribution and the observed data were transformed using the mean function.
Because we defined \sysname{} such that quantities in model samples and observed data should be on the same scale, this use falls outside the scope of our correctness objective (\ref{req:correctness}).
These observations were useful for identifying how such usage can potentially be useful in scenarios like hypothesis testing on certain statistics.
On the other hand, if comparability is not strictly ensured, additional actions on the visual front are necessary.
For instance, when transforming the observed data by standard deviation function, specific visual techniques should be used to clarify that the observed data is now represented on the variance scale.

\subsubsection{RQ2: Usability of the Grammar}

While going through the tutorial, participants commented on how they reasoned about \sysname{} specifications.
Two of the three participants (P1 and P3) commented on how \sysname{}'s specification of the data preparation step was a good abstraction of how they think and express data targets in model check visualizations.
P3 compared his experience with \texttt{bayesplot} to \sysname{}, saying ``\textit{[In \sysname{}, ] I only need to specify mu [for the model quantity] and mean [for the data transformation]. It is really good compared to that [bayesplot]}''.
P1 also found the naming convention and modular structure  resembling that of \texttt{ggplot2} of \sysname{} familiar to him, which made it intuitive to him -- ``\textit{I had a pretty good idea of what it was gonna do... [because] it's modular, and the names make sense. So it's pretty intuitive.}''

All the participants appreciated that \sysname{} allows them to specify model check visualizations in a high-level and systematical way.
P3 mentioned that high-level interfaces for the observed data and the model make it easy to think of model check visualizations. Referring to the specification, he said ``\textit{There is the data versus model. I do like the contrasting. I teach a Bayesian course in university and I can totally see myself like, teaching students how to use this. I like that. It's relatively easy to transform things or extract things.}'' 
That \sysname{} is built on a succinct specification with a few conceptual components was deemed effective by multiple participants:
``\textit{[The part I like about the grammar is that] it is built on these four or five foundational functions that built these plots}'' (P2).

However, participants also noted some negatives.
One of the biggest concerns was that despite how \sysname{}'s design separates components, they often remain entangled in some ways.  For example, when specifying the mark type as line+ribbon in \visualrepresentation{}, there must be a conditional variable on the x-axis.
In trying to run a \sysname{} specification component by component, P3 commented on this dependency between visual representation and conditional variable:
``\textit{The grammar s like [independent] layers added up. But it's not really [independent] layers. The visual representation layer can not work independently without the conditional variable layer.}'' 


\subsubsection{RQ3: Exploration}

Participants used different approaches in generating the new visualizations by applying \sysname{} to the models they brought.
Each participant adapted five specifications on average in the time they had.
P1 started by copying all the examples in the tutorial and changing the model input to his model.
After he had reviewed all examples, he chose to explore the argument that he had not yet encountered: \texttt{grouping\_sample} in the \visualrepresentation{} component.
He tried combining different mark types with HOPs.
P2 sought better alternatives starting from specific examples from the tutorial, trying to reason about what should be changed in applying them to his model.
For example, when looking at the raincloud example~\autoref{fig:expressiveness}A, he said ``I guess one thing that...It would be easier to encourage people to think about data if these were histograms instead of dots''.
He changed the \texttt{mc\_obs\_dots} to \texttt{mc\_obs\_histogram} to check that it was straightforward to change the mark.
P3 followed a more top-down approach.
He started by thinking of the general goals he had for the plots then tried to translate those concepts into the components of our grammar.
He first thought that he wanted to check the $\sigma$ parameter in the Gaussian distribution, so he specified \texttt{mc\_draw("sigma")} and \texttt{mc\_observation\_transformation(sd)}.
Then he tried to find a suitable visual representation for the model samples and the observed data.
He went through points, lines and line+ribbon mark types, choosing the line+ribbon as the final representation.

Although all the participants succeeded to explore new visualizations using \sysname{}, they did not always find the newly generated visualizations effective.
For example, P2 experimented with different mark types on the observed data. 
He discovered that marks utilizing smoothing methods failed to accurately represent the original data instances, potentially leading to misinterpretations.
The flexibility of \sysname{} also exposed some ineffective edge cases such as mismatched data and visual representations.
For example, during P3's exploration, he found that mark types for density estimates should not be applied to the observed data when it is aggregated into a single value such as the mean, although this case is not precluded by \sysname{}.
Nonetheless, that participants could quickly identify the effectiveness of the newly generated model check visualizations is a promising sign that \sysname{} could lead modelers to identify new, useful designs.

\section{Discussion}
We developed \sysname{} to explore the potential of using a small set of sensible abstractions to make model check visualizations easier to reason about and generate. 
In its current iteration, \sysname{} can be of immediate use in statistical workflows. 
In exploratory data analysis, \sysname{} can be used to explore provisional models used to drive understanding of features and heterogeneity in the data. 
In a Bayesian workflow, \sysname{} could be used to check the prior distribution before model fitting, and support comparisions of alternatives during fake data simulations~\cite{gabry2017visualization}.
In any statistical workflow, after the model is fitted, \sysname{} can be used to check predictive distributions, apply test quantities, and check model assumptions like outliers, distributional assumptions and heteroskadasticity. 

While the development of \sysname{} was inspired by Bayesian workflow where graphics have been more widely accepted as a crucial part of modeling workflow, one of our aspirations in developing \sysname{} is that such tools may be helpful in popularizing graphical model checks in a wider range of statistical workflows as well as statistics education. For example, standard materials for learning regression often focus on only a handful of canonical visualizations (residual plots, QQ plots). By formulating visual model checks as a more extensible design space, tools like \sysname{} may contribute to more effective modeling in practice by calling greater attention to the many subtle ways in which model predictions may align with or deviate from observed data.

Beyond its immediate practical value, developing and evaluating \sysname{} led to several observations about graphical model checking as a tool in statistical workflow.

First, our work highlights the \textbf{need to account for both the visual judgment mechanisms and data alignment required to facilitate comparisons of relevant quantities}.
Visualizations intended for comparison should not only facilitate visual judgments through effective comparative layouts, they should  also consider how to ensure comparability of quantities through data alignment.
\new{For instance, drawing meaningful conclusions by comparing the intercept parameter from a linear regression model with the observed outcome values is unlikely, regardless of whether they are superposed or juxtaposed. }
While our constraints in developing \sysname{} precluded some useful examples (e.g., \new{when the observed data is simulated and the intercept used for simulation known}), encouraging authors to think carefully about alignment is likely useful.
There may be some value in type systems for data scales, similar to formal guarantees in typed programming languages, that can help people align the quantities effectively and avoid comparing them on scales that do not match.

Second, a tension we encountered in developing \sysname{} points to \textbf{the value of extending API support to \textit{non-checkable} model quantities to support important steps in a statistical workflow}, such as setting and interpreting priors or displaying fitted parameter values.
According to our definition, a \textit{checkable} model quantity must have a transformation function that can align the observed data with it on a common scale.
However, for \textit{non-checkable model quantities}, no such transformation function exists, and we must turn to other design strategies.
When the goal is to make model checking accessible to a broader set of modelers beyond experienced statisticians, then it would be useful for future work to identify systematic ways of contextualizing abstract scales like parameter spaces. Doing so could make 
visualizations of
non-checkable quantities more concrete, similar to how a model check that requires checking against observed data makes model quantities easier to understand.
This could be achieved, e.g., through
analogies to known phenomena \cite{kim2017explaining} or reference markings illustrating the relationship between the abstract scales, like the explode-y graphics in \cite{yang2023subjective}.

Third, widely applicable model checking visualization tools call for \textbf{appropriate software abstractions to overcome heterogeneity in model outputs, syntax, and checkable quantities}, so that model check visualization tools like \sysname{} can be widely applied. 
In the implementation of \sysname{}, we aimed to support different model objects and checkable quantities by combining different model extracting libraries, e.g., \texttt{tidybayes} \cite{tidybayes} and \texttt{insight} \cite{insight}.
Other model inspection R packages like \texttt{marginaleffects} \cite{marginaleffects} and \texttt{easystats} \cite{easystats} 
take steps to write a layer of abstraction that accommodates all of these variations without requiring intervention by the user, increasing the interoperability of tools for modelers within the R ecosystem.
Such efforts are a natural next step for extending \sysname{}, though complete interoperability may not be possible as a result of the continually changing landscape of modeling libraries.
\new{
Our approach in \sysname{} is also compatible with model comparison tools that statisticians often use, such as leave-one-out (LOO) cross-validation and multiverse analysis, but these applications require extensions that accommodate multiple model objects.
For example, users can extend \texttt{mcplot} to take a list of models as input.
As long as each model in this list has the same model quantities, it is straightforward to apply the visual representations and comparative layout in \sysname{} to them.
It is also worth extending operations to the predictions of these models after the \textit{sample} stage in the compiler to support more complex abstractions in the statistical tools, e.g., model comparison in LOO cross-validation.}

Along the lines of a proposal for more explicit support for model checking in visual analysis tools~\cite{Hullman2021Designing}, \sysname{} can be incorporated in interactive analysis tools to support the generation of model checks in exploratory data analysis, similar to the design approach by Kale et al. in EVM~\cite{kale2023evm}.
Adapting \sysname{} for this purpose would allow the developer to avoid the diffuseness and viscosity of the notation in lower-level visualization grammars and focus squarely on the main substantive components of a model check.


Better tools for graphical model checking can also improve the development of predictive models through better designed VIS4ML systems, where principled processes are still in development.
Components of \sysname{} can be taken as a foundation for the development of such tools, and extended to support high dimensional datapoints like images.
Understanding what makes a good model check with (parametric) statistical models may not be equivalent to what make a good model check with machine learning models, motivating empirical work.

\subsection{Limitations}

While we propose four necessary design objectives of \sysname{} 
based on the practice and theory of model checking in the statistics and visualization literature, we cannot guarantee that these are the sufficient for a good model check visualization grammar.
For example, a model check grammar could benefit by facilitating the specification of common 
interaction techniques in visualizations 
(e.g., brushing and linking)
such that users can quickly find patterns that reflect important types of misfit~\cite{kale2023evm}.
Other missing design objectives may include expressing sampling primitives and facilitating model comparison workflows.
Future work is needed to adjust the design objectives based on the target scenario.

\sysname{} specifications are also limited to position encodings (i.e., x-axis, y-axis, rows, and columns), which rules out visualization techniques that display data without axes, e.g., maps or some space-filling techniques.
Allowing display without standard axis encodings may require another abstract layer in \sysname{} to map the data onto the display space, e.g., mapping a country variable to a spatial region on the map. 
\acknowledgments{%
	We thank our anonymous reviewers for their helpful suggestions. Jessica Hullman thanks NSF \#2211939 for supporting this work.%
}

\bibliographystyle{abbrv-doi-hyperref}

\bibliography{model_check}

\appendix

\section{Implementation in R}
\label{sec:app_impl}



We implement \sysname{} in the R language as a package, provided in Supplemental Materials.
The specification of \sysname{} follows conventions popularized by the widely-used \texttt{ggplot2} package and outputs a \texttt{ggplot} specification to generate the model check plot.
To start specifying a model check plot, the user issues a statement called \texttt{mcplot()} and then follows it with statements specifying the four requisite components of \sysname{}: \texttt{mc\_draw()} for \samplingspecification{}; \texttt{mc\_observation\_transformation()} for \datatransformation{}; \texttt{mc\_model\_*()} and \texttt{mc\_obs\_*()} for \visualrepresentation (where \texttt{*} stands for different mark types, e.g., \texttt{mc\_model\_slab()} for using \textit{slab} for model predictions); and \texttt{mc\_layout\_*()} for \comparativelayout (where \texttt{*} stands for different layers, e.g., \texttt{mc\_model\_superposition()} for using \textit{superposition}).
The implementation of \sysname{} also supports other supplementary components, e.g., \texttt{mc\_condition\_on()} to define the conditional variable used in the check and \texttt{mc\_gglayer()} to append a \texttt{ggplot2} layer to the output specification.


The \sysname{} compiler generates the output \texttt{ggplot2} specification in four stages: 
\textit{sample} from the model-defined distribution producing a tidy data frame, \textit{transform} the observed data, \textit{translate} the visual representation components into \texttt{ggplot2} geometry layers, and finally \textit{construct} the comparative layout to generate the final \texttt{ggplot2} objects.

In the first stage, the compiler extract the samples defined in the component \texttt{mc\_draw()}.
The compiler is designed to support as many model classes as possible.
For Bayesian models such as \texttt{brmsfit} and \texttt{stanreg}, the compiler translates the specification of \samplingspecification{} using functions in the \texttt{tidybayes}~\cite{tidybayes} package and supports model quantities for any combinations between link functions and distributional parameters.
For non-Bayesian models such as \texttt{gam} and \texttt{lm}, the compiler translates the specification of \samplingspecification{} using functions in the \texttt{insight}~\cite{insight} package, and supports model quantities such as link-scale predictions and response scale predictions (either the expectation of predictions or the predictions themselves).
Missing arguments are inferred using sensible defaults.
For example, by default the compiler uses the fitted data as the predictor values if it is not specified.
The compiler next transforms the data based on the specification of \datatransformation{}.

After completing data preparation, the compiler translates the components \texttt{mc\_model\_*} and \texttt{mc\_obs\_*} into \texttt{ggplot2} components, applying the sampled data and creating aesthetics for the \texttt{ggplot2} components.
The compiler creates the aesthetics from the response variable in the model and the conditional variable specified by \texttt{mc\_condition\_on}.
We also implement recommender components, \texttt{mc\_model\_auto} and \texttt{mc\_obs\_auto}, for \visualrepresentation{}, which use a default visual representation based on the type of response variable and conditional variable.
The compiler generates two groups of \texttt{ggplot2} components from the visual representation components of \sysname{}, one for the model predictions and one for the observed data.

Finally, the compiler constructs the \texttt{ggplot2} specification for the model check visualization.
It obtains the two groups of \texttt{ggplot2} components for model predictions and observed data, then merges them based on \texttt{mc\_layout\_*} in the \sysname{} specification.
For superposition, the compiler generates one \texttt{ggplot2} object to overlay the two groups of \texttt{ggplot2} components in the same axes.
For juxtaposition, the compiler generates two \texttt{ggplot2} objects for each group and organizes the layout to juxtapose them.
For explicit encoding, the compiler computes the transformation that the user specified to encode the difference between model predictions and observed data.
The compiler predefines some common used explicit encoding layouts, namely residual plot, Q-Q plot and worm plot, to avoid requiring users to define their own transform functions every time.

\end{document}